\hsize 6.25truein \vsize 9.8truein \topskip 1truecm \voffset -.2truein
\hoffset 0.5cm
%
\newskip\footnotebaselineskip
\footnotebaselineskip=\baselineskip 
\def\footnotefont{\rm} 
\catcode`\@=11 
\def\footnote#1{\let \@sf \empty
 \ifhmode \edef \@sf {\spacefactor \the \spacefactor }\/\fi
  $^{#1}$\@sf \vfootnote{#1}
}
\def\vfootnote#1{
 \insert \footins \bgroup
 \interlinepenalty \interfootnotelinepenalty
 \splittopskip \ht \strutbox
 \splitmaxdepth \dp \strutbox \floatingpenalty 20000
 \leftskip 0pt \rightskip 0pt \spaceskip 0pt \xspaceskip 0pt
 \baselineskip \footnotebaselineskip \footnotefont
 \textindent {$#1$}\footstrut \futurelet \next \fo@t
}
\catcode`@=12 
%
\newskip\chapheadskip
\chapheadskip = 24pt plus 8pt minus 8pt
\def\mysection#1{\par\goodbreak\vskip0.6\chapheadskip
\global\advance\count2 by 1%
\dimen255=\pagegoal\advance\dimen255 by -\pagetotal
\ifdim\dimen255<1truein \message{space left \the\dimen255:
page ejected: }\vfil\eject\message{(\the\count1.\the\count2\ #1)}%
\else \message{(\the\count1.\the\count2\ #1)} \fi
\ifnum\count1=0%
\leftline{\sectname\quad\the\count2. #1}\smallskip\vskip-\parskip \noindent
\else
\leftline{\sectname\quad\the\count1.\the\count2. #1}\smallskip\vskip -\parskip
\noindent\fi}%
\def\mypart#1.{\par\goodbreak\vskip0.6\chapheadskip
 \dimen255=\pagegoal\advance\dimen255 by -\pagetotal
 \ifdim\dimen255<1truein \message{space left \the\dimen255:
 page ejected: }\vfil\eject\message{(#1)}
 \else\message{(#1)}\fi
 \leftline{\bigbold\quad#1}\bigskip\noindent}
%
\newcount\refno \newif\ifref \reffalse \newif\ifcom
\def\bref{\relax\reftrue\comfalse[}
\def\eref{\relax\reffalse\comfalse]}
\def\newref{\global\refno=0\relax\gdef\reffage{}} \newref
\def\donewref{\newref}

\def\ref#1{\global\advance\refno by 1\relax\ifref \ifcom,\fi \the\refno\relax
\comtrue\relax\else[\the\refno]\fi
 \xdef\reffage{\reffage\hang\textindent{[\the\refno]}%
\hangindent\parindent \hangafter=1 #1\par}}
\def\moreref#1{\xdef\reffage{\reffage\vskip-\refskip \hang#1\par}}
\def\keepref#1{\xdef#1{\the\refno}}
\def\rref#1#2{\ref{#2}\keepref#1}
\def\oref#1{\ifref\ifcom,\fi#1\comtrue\else[#1]\fi}
\newskip\refskip
\refskip=12pt plus 1pt minus 1pt
\def\references{
 \mypart{References}.%
 {\pretolerance=10000 \raggedright \baselineskip=\normalbaselineskip
  \parskip=\refskip \reffage}
 \donewref 
}
%
\newcount\mynum
\newcount\mypartnum
\def\newnum{\global\mynum=0 \global\mypartnum=96\relax}
\newnum
\def\num{
 \global\mypartnum=96
 \global\advance\mynum by1
 \xdef\cnum{(\ifnum\count1=0 \else\the\count1.\fi
  \the\mynum)}
 \cnum\relax
}
\def\numpart{
 \ifnum\mypartnum=96 \global\advance\mynum by1\fi
 \global\advance\mypartnum by1
 \xdef\cnum{(\ifnum\count1=0 \else\the\count1.\fi
  \the\mynum)}
 (\ifnum\count1=0 \else\the\count1.\fi
  \the\mynum{\rm \char\mypartnum})\relax}
\def\rnum#1{\num\xdef#1{\cnum}}

\def\eq#1{eq.~#1}
%
\skip\footins=20pt plus 6pt minus 6pt 
\footnotebaselineskip=12pt 
\def\footnotefont{\ninerm} 
\font\ninerm cmr9 
\font\sectname cmsl10 scaled \magstep1	
\font\stepone cmr10 scaled \magstep1
\font\bigbold cmbx10 scaled \magstep1
\font\biggerbold cmbx10 scaled \magstep2
\newnum \newref
\baselineskip 15pt

\line{SWAT/13\hfil hep-lat/9310019 \hfil October 1993}
\bigskip
\centerline{\biggerbold The running coupling from lattice gauge theory}
\medskip
\centerline{\bigbold Talk at the Workshop on Field-Theoretical
Aspects of Particle Physics,}
\smallskip
\centerline{\bigbold Kyffh\"auser, Germany, September $13^{\rm th}$ to
$17^{\rm th}$ 1993}
\bigskip
\centerline{P.W. Stephenson}
\smallskip
\centerline{Department of Applied Mathematics and Theoretical Physics,
University of Liverpool}
\centerline{and}
\centerline{Department of Physics, University College of Swansea}
\centerline{Singleton Park, Swansea, SA2 8PP, UK.\footnote{*}{Present
address.   Email:  P.Stephenson@swan.ac.uk or PWS@UKACRL (BITNET).}}

\bigskip

\centerline{Abstract}

{\narrower\noindent I discuss some calculations of the running
coupling in SU($N$) gauge theories from lattice simulations, centering
on the work of the UKQCD collaboration.  This talk is introductory in
nature; full details have been published elsewhere.
\par}

\bigskip
\mysection{Introduction}%
At the moment, lattice simulations are the most popular way of
extracting truly non-perturbative results from quantum field theories.
Their commonest uses are, quite naturally, calculations with some
direct experimental relevance, such as the spectroscopy of hadrons and the
calculation of matrix elements.

However, we are quite at liberty to examine more basic aspects of the
discrete theory, in order to assure ourselves that the results are as
we expect.  (It would be more interesting if they were not, but in QCD
this is increasingly unlikely.)  In this talk I shall describe a
(successful) attempt to look at the behaviour of the running coupling
of the field theory.  This allows us to make direct contact with the
standard formalism of perturbation theory and the renormalisation
group.

Phenomenologically, the most interesting field theory is quantum
chromodynamics (QCD), with its strong interaction and significant
scale-dependence over the region where the physics is most
interesting.  The proper lattice formalism is that of SU(3) gauge
theory with {\bf dynamical} quarks, incorporating Feynman diagrams
with internal quark loops.  This is quite simply too difficult for us
with our present-day technology and computer resources.  Even
generating an equilibrated lattice (one on which samples of the fields
are guaranteed to be representative of the true vacuum with the right
coupling) is very costly and can take many months even on the largest
machines.

However, simulations in the {\bf quenched} approximation, the theory
without dynamical quarks but retaining fully non-perturbative gluonic
contributions, are proving more tractable and we are seeing results
which in many cases agree with experiment at the level of 10\% or so.
One should of course be careful; at some stage the effect of quenching
is likely to dominate our errors and in some calculations to change the
nature of the physics completely.  We are not yet at the stage where we
can determine the limits of the quenched approximation and in general
the effect is not predicted.  Preliminary indications are that where
light quarks are unimportant the dominant effect is simply a uniform
rescaling of the results.

This need not worry us if we are engaged in an exercise in field theory.
In fact, in that case we can make one further simplification and use the
gauge group SU(2) instead of SU(3), resulting in roughly an order of
magnitude reduction in the computing effort required for a similar
standard of results.  We shall also see that the results for the two
groups are qualitatively very similar.

This talk deals for the most part in the pure gauge theory, which does
not involve fermions at any stage.  This is not a further
approximation beyond quenching:  it simply means we are looking at the
gluonic sector.  In fact, the pure gauge theory is a true field theory
(to the best of our knowledge) while the quenched fermionic theory is
not since it involves an unnatural treatment of quarks on a background
gauge field with which they do not interact properly.

The technology for producing pure-gauge results from lattice simulations
is now quite well advanced.  Using improved operators we can extract
quite accurate numbers for, among other things, the interquark potential
which will be our probe here.

The method outlined in the major part of the talk first appeared in
\rref\cmalpha{C.~Michael, Phys.\ Lett.\ B283 (1992) 103}. Full details
of the calculations by the UK QCD Grand Challenge (UKQCD) using this
method are given in ref.~\rref\ukqcdtwo{UKQCD Collaboration, Nucl.\
Phys.\ B394 (1993) 509} (for SU(2)) and ref.~\rref\ukqcdthree{UKQCD
Collaboration, Phys.\ Lett.\ B394 (1993) 509}.

\mysection{The running coupling and asymptotic freedom}%
In QCD, or any quantum field theory, the physically-meaningful
coupling --- the one which relates directly to the strength of the
interaction --- is the renormalised value.  This is commonly referred
to as the {\bf running} coupling, as the process of renormalisation
makes the coupling apparently change its value in low-order
perturbation theory.  One way of looking at this new quantity is that it
parametrises our ignorance about what is happening at very high
energies when QCD is no longer valid, replacing it with a cut-off and
an effective theory.

\def\LQCD{\Lambda_{\rm QCD}}

In four dimensions, where the coupling is dimensionless, we see the
rather surprising phenomenon known as dimensional transmutation
whereby the mechanics of renormalisation introduces a fixed scale,
namely the lambda parameter, $\LQCD$.  The ``interesting'' physics of
QCD, by which I mean the battleground of different causes and effects,
happens in processes involving momenta $q \sim \LQCD$.  The second
important consequence of dimensional transmutation is that the
renormalised coupling is itself scale-dependent, $g_{\rm phys} =
g_{\rm phys}(q)$.  As we shall see later on, the connection between
$g_{\rm phys}$ (I shall drop the ``phys'' suffix) and the bare
coupling $g_0$ which appears in the original Lagrangian can be a bit
obscure.  Actually, this is true not just on the lattice: it is, after
all, why we need renormalisation in the first place.

The SU($N$) theories have the feature that as the momentum scale is
increased, the physical coupling decreases ({\bf asymptotic freedom}).
This means that in the limit of large momentum the theory becomes
perturbative.  That is how we shall connect the non-perturbative
lattice results with analytic calculations in perturbation theory.

In SU($N$) to two loops the running coupling is given by $$\alpha(q)
\equiv {4\pi\over g^2} = {1\over 4\pi\left(b_0\log (q/\Lambda)^2 +
(b_1/b_0)\log\log (q/\Lambda)^2\right)}\eqno\num $$ where $b_0$ and
$b_1$ depend only on $N$.  This is for the pure gauge theory with no
quarks; adding a few species of quarks does not change the expression
qualitatively, though asymptotic freedom is weakened and eventually
(with 17 quarks in SU(3)) the sign of $b_0$ and the character of the
theory change.  

Instead of a momentum scale, one can express the results in terms of
an appropriate length scale $R\sim q^{-1}$.  This is more appropriate
to a static configuration like the one we use for extracting
potentials.  We need to change our renormalisation scheme to do this;
the scheme appropriate to potentials uses the separation $R$ between
the quark and antiquark, as described in the next paragraph, as the
scale parameter.  It turns out that this is close to $\overline{\rm
MS}$ in the sense that the $\Lambda$-parameters are related by a small
factor; this is not true of the usual scheme for lattice
regularisation.

In the perturbative limit, the force between a static (infinitely
massive) quark and antiquark of opposite colour (so the pair is
colourless) a distance $R$ apart is Coulombic:
$$-F_{\rm Coulomb}={dV\over dR}=C_f{\alpha(R)\over R^2}\eqno\num$$
($C_f$ is a combinatoric factor containing $N$; it derives from the
quadratic Casimir operator).  Hence if we form a dimensionless quantity
$$\alpha_{\rm eff} \equiv -{F_{\rm calc}R^2\over C_f} \eqno\rnum\aeff$$
a lattice calculation of $F_{\rm calc}$ at small $R$ is enough to tell
us the running coupling.
We use the force instead of the potential as it removes an
uninteresting constant.  On the lattice, the force is extracted from
the potential by a finite difference; this actually introduces
negligible extra errors.  The significant thing about the lattice
calculation is that, although we are trying to show agreement with the
perturbative analysis, the calculation is at every step fully
non-perturbative.

\mysection{Lattices and scales}%
Our lattice formulation is the standard one of Wilson.  The gauge fields
live on the {\bf links} of a hypercubic lattice in Euclidean
space-time, that is the lines joining nearest-neighbour points of the
simple cubic structure like a stack of wire-framed cubes.  Each
link has an associated matrix which is an element of the gauge group.
Line integrals between points are replaced by simply matrix
multiplication of the appropriate links; taking the trace of such a
product which corresponds to a closed loop produces a gauge-invariant
object, the {\bf Wilson loop}.  These loops are essentially the only
gauge-invariant quantity in the pure  theory; the quarks, if
present, would live at the sites of the lattice and act as sources and
sinks of colour.

The action for the theory is a sum over all {\bf plaquettes}: the
plaquette is the smallest possible Wilson loop consisting of four links
about an elementary square of the lattice.  This preserves exact gauge
invariance and in the continuum limit it goes over smoothly to the usual
continuum action.

The scale is fixed by the {\bf lattice spacing} $a$, the length
of one link.  This is determined by the bare coupling $g_0$ which we
choose for our interaction (in fact, we usually work with
$\beta=2N/g_0^2$, which appears as the multiplier of the sum of
plaquettes in the action).  The relationship between $a$ at different
values of the coupling is determined by the renormalisation group
beta-function (two completely unrelated uses of the symbol beta,
unfortunately); because of asymptotic freedom, a small $g_0$
corresponds to a small lattice spacing.  Decreasing $g_0$ therefore
makes the lattice less and less important and takes us nearer the real
world where $a\Lambda$ is small, or in other words the scale $\Lambda$ is
very much less than the momentum space cut-off.

\midinsert
\vskip 4in
\includegraphics{wloop.eps}
\smallskip
\leftline{\bigbold Figure~1\stepone:  The Wilson loop on the lattice.}
\medskip
\endinsert

One can think of a rectangular Wilson loop (as in figure~1) which
extends $R$ in a spatial direction and $T$ in the time direction in the
following way.  A source and a sink of colour in the fundamental
representation of the gauge group, both infinitely massive, are created
instantaneously a distance $R$ apart.  They propagate in this fashion
for time $T$ and then are instantaneously annihilated.  We refer to the
colour sources as a (static) quark-antiquark pair even though the
calculation does not involve fermions at any stage.  We need some
experimental input to determine $a$ (or --- which is equivalent in
theory if not always in practice --- $\LQCD$) in physical units.

\mysection{Measuring the quark-antiquark potential}%
To extract the potential between the two, we simply need the expectation
value of the loops.  This is related to the Euclidean propagator
$C(T) \sim \sum_ic_i\exp(-V_iT)$ for the various energy levels $V_i$
of the quark-antiquark system.  At large $T$ the main contribution is
from the ground state and the ratio $C(T)/C(T-1)$ tells us the potential
$V\!\equiv V_0$.  Unfortunately the statistical errors increase with $T$,
also the decay of the correlators is faster than typically found in (for
example) calculations of light hadron masses, so choosing a suitable
value of $T$ is something of an art form and introduces significant
systematic errors.

We should note two other major sources of systematic error in our
calculations, again deriving from limitations on computing resources.
First of all the lattice size $L$ is finite as we can only fit $L/a$
sites on a side of our lattice.  If this is too small the fields will
feel the effects of the boundaries of the lattice, being squashed into
the box.  In our case we have chosen sizes (from previous experience)
such that this is not expected to be a problem.

Secondly, $a$ is finite, generating cut-off effects.  This is
significant for us, since we are attempting to probe the region at
small $R$ where perturbation theory is expected to become valid: our
results are for quark separations of only a few lattice spacings.  We
have used a one-parameter fit to smooth out the bumpiness caused by
finite $a$; the smoothness of the result, together with the agreement
between different lattice spacings as described below and shown in
figure~2, assures us that this has worked.

Our Monte Carlo simulation generates different samples of the QCD
vacuum for us to measure by {\bf local updating}:  changing one link at
at a time until the set of all links is sufficiently different from the
previous sample.  The process involves subjecting each link to
two different procedures.  The first is a ``heatbath'', closely
analogous to the same concept in statistical mechanics, in which the
link is made aware of the coupling (corresponding to the temperature) of
the links around it.  This includes the element of randomness that
drives our stochastic process, like the random kinetic motion in a
real heatbath.

The second procedure is ``over-relaxation''.  This uses the fact that
the SU(2) group manifold is a sphere and that we can flip the gauge
element for any link about the (scaled) gauge element representing the
combined effect of the other plaquettes in which the chosen link
appears without changing the action.  Hence we can do this as often as
we like without affecting the statistical-mechanical properties of the
simulation.

In both procedures, each link of the lattice is updated in turn
throughout the whole lattice (one {\bf sweep}).  The SU(3) simulation
is similar: for each link we perform a heatbath in each of the three
possible SU(2) subgroups in turn, and similarly for over-relaxation.
Typically we perform four over-relaxation sweeps for every heatbath
sweep; some other groups perform more, but our heatbath code is well
optimised so that the extra computer time over the over-relaxation
code (which is simpler) is fairly low.  This appears to be the most
efficient way of generating distinct configurations at present.

The sets of data we have used are in each case separated by several
hundred sweeps (largely for logistical reasons) and statistical
correlations between different sets are found to be completely
negligible.

Having extracted values for the potential we should like to decide
whether results on lattices at different inverse couplings $\beta_1$,
$\beta_2$ give equivalent results --- in other words, whether the ratios
of measured quantities (say, masses) $m_X(\beta_1)/m_X(\beta_2)$ are the
same for all possible measurements $X$.  If this happens, the results
are said to {\bf scale} and we know that any dependence on
$a$ in the expansion of the ratio has disappeared.  (However, we cannot
necessarily make the stronger statement that the individual quantities
$m_X$ have an expansion which behaves according to the RG beta
function in low order.  This requirement --- {\bf asymptotic scaling} --- is
mentioned later.)

\topinsert
{\vskip 4in
\includegraphics{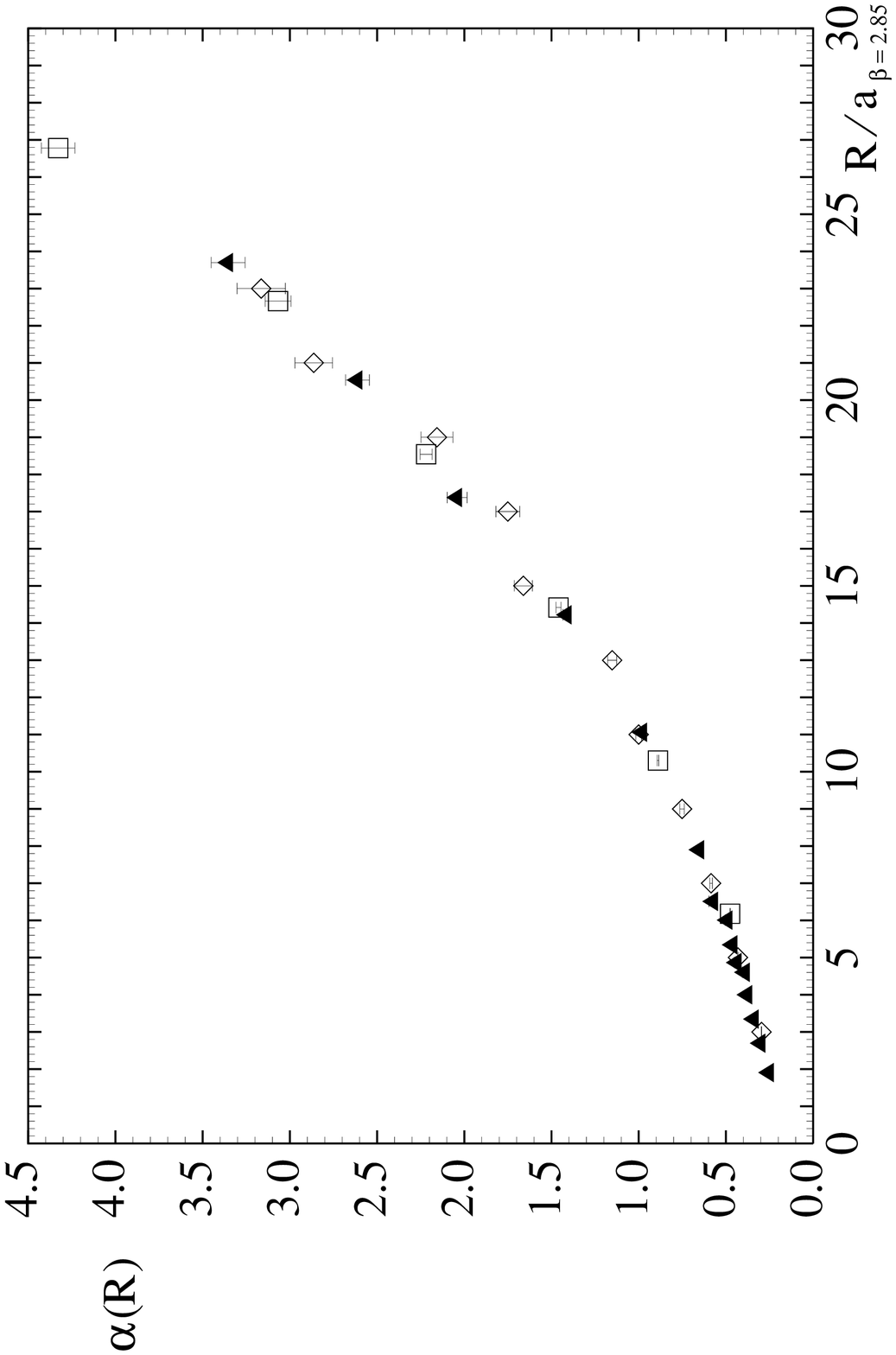}
\noindent \bigbold Figure~2\stepone:  The scaling properties of the potential
in
SU(2).  Diamonds are $\beta=2.85$ data, triangles $\beta=2.7$ and
squares $\beta=2.4$.  The $R$ axis (only) is adjusted for the latter two.}
\endinsert

One way of showing this from the potential data is by forming
``$\alpha_{\rm eff}$'' as in \eq\aeff.  We do not require at this
point that this $\alpha_{\rm eff}$ should be the true running
coupling; we are simply using it as a convenient dimensionless
physical quantity.  If scaling holds we should be able to plot this
against the separation $R$ and achieve a single curve simply by
rescaling the $R$ axis as appropriate.  Figure~2 shows this for three
different couplings in SU(2): $\beta = 4/g_0^2 =
2.4$~\ref{S.J.~Perantonis, A.~Huntley and C.~Michael. Nucl.\ Phys.\
B326 (1989) 544}, $2.7$~\oref\cmalpha and 2.85~\oref\ukqcdtwo.  The
scaling factor in $R$ shows that the lattice spacing is some four
times smaller at $2.85$ than at $2.4$.  This is a strong indication
that we have control over all finite lattice spacing effects --- any
scaling violation in the potential measurement should show up clearly
in this plot.

\mysection{The running coupling}%
We now proceed to the running coupling itself (following the method of
ref.~\oref\cmalpha).  The potential
shown in figure~2 was over a wide range of scales which includes a
linearly-rising potential $V \sim KR$ that dominates at
large $R$ (we confirm this by suitable fits to the data; the value of
$K$ is well-determined).  We concentrate on our smallest lattice
spacing, with $\beta=2.85$ and 48 lattice sites in each spatial
direction, for small separations.  As the lattice spacing is
unphysical, we set the scale instead by using our measured value of
$K$ on the same lattice.  This result is shown in figure~3 (the points
with error bars).  We show the corresponding analysis for SU(3) in
figure~4; in this case $\beta = 6.5$ and the lattice has 36 sites in
each spatial direction.

\pageinsert
{
\null\vfil
\includegraphics{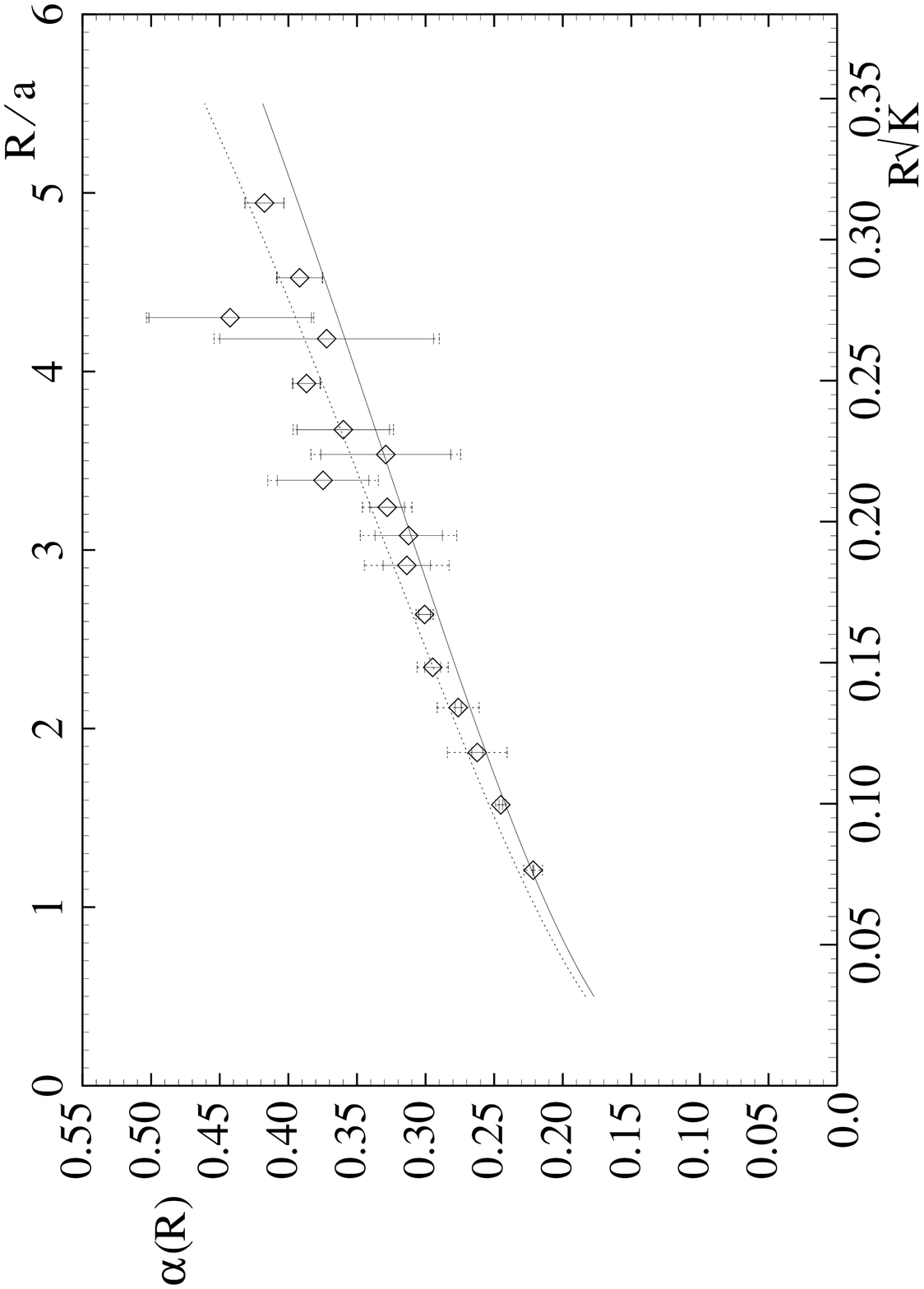}
\noindent \bigbold Figure 3\stepone: The running coupling in pure
SU(2) lattice gauge theory.  All results are at $\beta=4/g_0^2=2.85$.
Error bars are both statistical and systematic; the latter are dotted.
The upper scale on the horizontal axis shows the separation in lattice
spacings; the lower scale relates it to the string tension.  The upper
and lower lines are two-loop perturbative predictions with
$a\Lambda=0.044$ and 0.038 respectively.
\vfil
\includegraphics{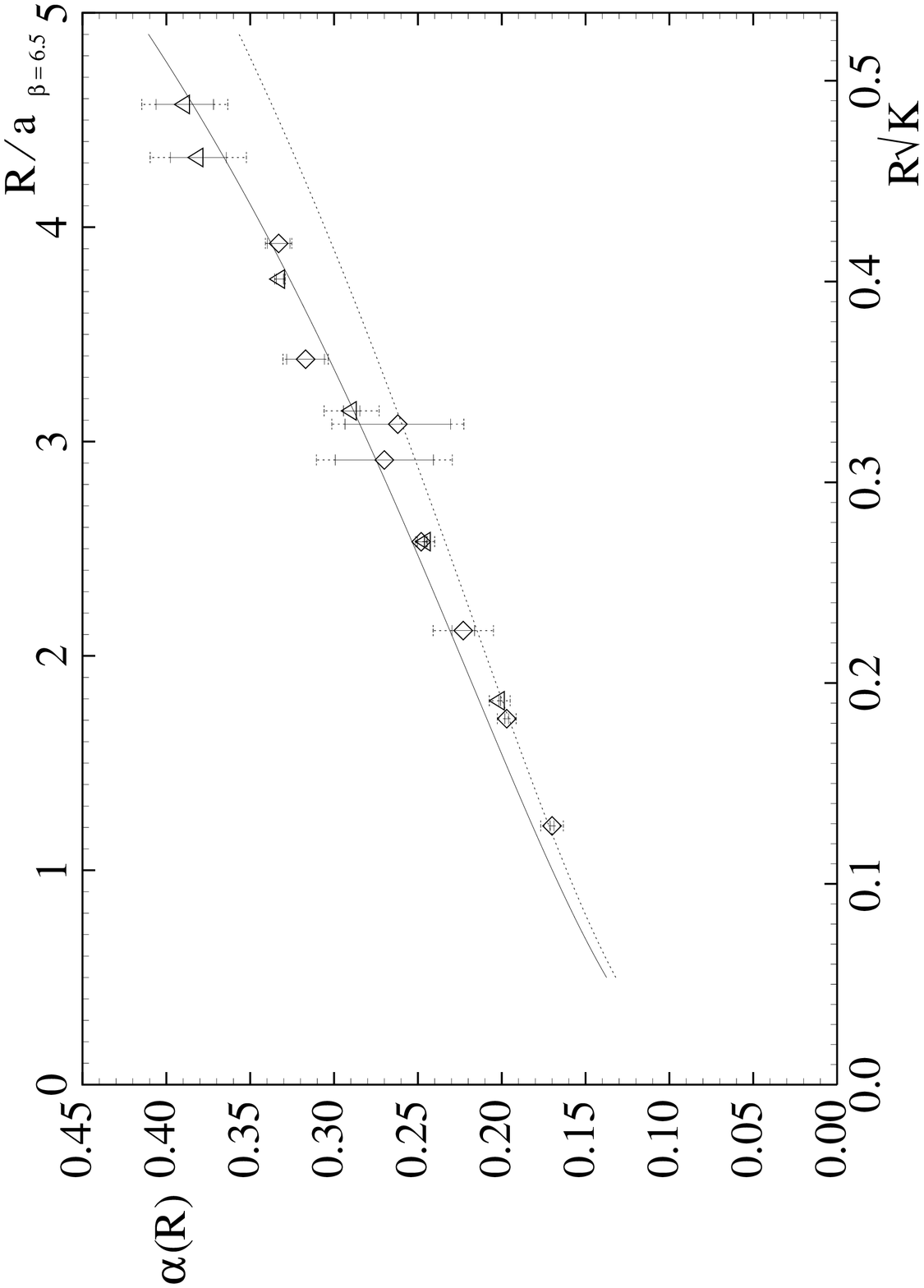}
\noindent \bigbold Figure 4\stepone: The running coupling in pure
SU(3) lattice gauge theory.  The plot is similar to figure 3.  Here
diamonds are data at $\beta=6.5$~\oref\ukqcdthree and triangles at
$6.2$~\ref{UKQCD Collaboration, Phys.\ Lett.\ B284 (1992) 377}; the
upper scale on the horizontal axis is only appropriate to the former.
The upper and lower lines in this case are for $a\Lambda=0.070$ and
0.060 respectively.  }
\endinsert

We should like to compare this with analytic results.  We do this by
choosing a value of $a\Lambda$ that fits our results.  The lines show
the running coupling from perturbation theory to two loops at the
largest and smallest values of $a\Lambda$ which seem consistent (this
was done by eye).  When this is done our lattice effective coupling
seems to agree well with the perturbative expressions (note that to
the right of the diagram non-perturbative effects are beginning to
enter).  In other words, we are seeing real perturbative field theory
from non-perturbative calculations on the lattice.

We can use our chosen value of $a\Lambda$ to extract $\Lambda$
provided we can express $a$ in physical units.  To do this we turn
back to $K$.  In SU(3), we can equate this with the ``string tension''
in the Regge picture of hadrons as a quark-antiquark pair connected by a
tube of relativistic glue.  (There is no more formal justification for
this; however, there is no good reason to suppose this is a bad way to
set the scale either.)  In this picture $\sqrt{K}=440$ MeV.  From
this, we extract the $\Lambda$ parameter in the renormalisation scheme
appropriate to the potential picture.  This is a small multiplicative
factor away from the more familiar $\Lambda_{\overline{MS}}$; we
deduce that $\Lambda_{\overline{MS}}=256\pm20$ MeV.  However,
remember this is in SU(3) without dynamical quarks.  We have not
attempted to correct for this; we do not believe that in our case the
systematic errors are sufficiently under control.

We can also use $K$ to give more meaning to the horizontal scale.
With the same value 440 MeV, a 5 GeV momentum scale corresponds to
$R\sqrt{K}\sim0.9$ on the lower scale of figure 4, so $\alpha \sim
0.16$.  This is very rough because apart from all the other errors I
have not bothered to do a proper conversion from the $R$ scheme to
$\overline{MS}$, though the difference is small; the physics is in the
running itself, not necessarily the actual values we extract.

\mysection{Other calculations}%
The same calculations have been performed in SU(3) by Bali and
Schilling \rref\bali{G.S.~Bali and K.~Schilling, Phys.\ Rev.\ D47 (1993)
661}; their results are very similar to ours, although they have more data.
Among their lattices are some with smaller physical extent than ours;
this does not change the running coupling behaviour, so it seems that
the small distance physics is not strongly affected by finite size
effects.  This was not {\it a priori} obvious, though it is not too
surprising a result if one believes in the separation of physics at
different scales.


Other calculations (ref.~\ref{A.~El-Khadra, G.~Hockney, A.~Kronfeld
and P.~Mackenzie, Phys.\ Rev.\ Lett.\ 69 (1992) 729}; see also reviews
in ref.~\rref\leplat{G.P.~Lepage, Nucl.\ Phys.\ B (Proc.\ Suppl.) 26 (1992)
45}\moreref{A.S.~Kronfeld and P.B.~Mackenzie, ``Progress in QCD using
lattice gauge theory'', preprint \hbox{hep-ph/9303305}, to appear in
Ann.\ Rev.\ Nucl.\ Part.\ Sci.}) for $\alpha$ have been performed
using the charmonium potential (any such colourless, heavy
quark-antiquark system will do as the results are nearly
mass-independent).  Here too the potential is expected to be in the
perturbative region.  In this calculation the scale does not appear in
an explicit non-perturbative fashion (as in our calculations) and one
has to make sure the scale is chosen appropriately, as emphasised by
Stan Brodsky at this workshop, and that the problems with perturbation
theory mentioned in the next section are correctly handled.  These
authors have chosen to correct for the effect of quenching (amongst
other systematic effects), that is to predict the effect of four light
quarks using the renormalisation group; this increases the value of
$\alpha$ at 5 GeV from $0.140\pm0.004$ to $0.170\pm0.010$.  Of course
it can then be run in the same way to any interesting scale.

\mysection{Lattices and perturbation theory}%
What might appear more surprising is that we have such good agreement
with perturbation theory at all.  The na\"\i ve way of comparing with
perturbative results is to insert our coupling $g_0$ into perturbative
expressions (such as the beta function) and see what comes out; in the
asymptotic scaling limit we would find agreement.  In practice
this is not seen. For example, the beta function produces a ratio of
lattice spacings between $a(\beta=2.7)/a(\beta=2.85)$ which is very different
from our value (by almost $20\%$).

The problem arises from the use of the bare coupling.  All our
previous analysis of the running coupling was done without any
perturbative input; $\beta$ was used only to label our different
lattices and so $g_0$ never appeared in the calculations.  There is
clearly a problem with perturbation theory on the lattice which does
not appear if one restricts oneself to physical quantities as we have
done.

This problem has been elucidated during the last couple of years by
Lepage and Mackenzie~\ref{G.P.~Lepage and P.B.~Mackenzie, Phys.\ Rev.\
D48 (1993) 2250}.  The lattice regularisation is somewhat
unnatural for a field theory; large constants from tadpole diagrams
are introduced into perturbative series which become only painfully
convergent.  Provided a physical coupling is used there is no problem.
Lepage and Mackenzie give methods for improving perturbative
calculations along these lines, using some non-perturbative input such
as the average plaquette as a renormalisation factor.

A further result of this is that attempts to extrapolate to zero
lattice spacing (i.e.\ zero bare coupling) by some groups
\bref\oref\bali\rref\fhk{J.~Fingberg, U.~Heller and F.~Karsch, Nucl.\
Phys.\ B392 (1993) 493}\eref\ has required more sophistication.  One
method of proceeding is to define an improved coupling with real physics
in it.  The ``$\beta_E$'' effective coupling scheme, descended
originally from a suggestion of Parisi \bref\ref{G.~Parisi, Proceedings
of the XXth International Conference on High Energy Physics, Madison,
1980, eds.\ L.~Durand and L.G.~Pondrom, American Institute of Physics,
New York (1981) 1531}\oref\fhk\eref, is one way of doing this; in fact,
it is very much in the spirit of the Lepage-Mackenzie programme.
Any physically quantity (in this case the action, which is easily
measured on the lattice --- in fact it very nearly emerges as a
by-product of the way links are updated) may be expressed as a power
series in the coupling.  $$\langle S_{\rm plaq} \rangle =
{c_1\over\beta} + {c_2\over\beta^2} + {c_3\over\beta^3} + \cdots
\eqno\num$$ This series may have a poor convergence.  However, one can
truncate at some low order and invert it, expressing the coupling as
some function of the action, and use this truncated expression to define
an effective coupling: $$
\beta_E^{(1)}\equiv {c_1\over S}\eqno\num$$ (this is the first order
$\beta_E$ scheme; one can truncate at higher order).  By missing out
the higher terms, one hopes the new coupling is more appropriate to
use in low order, having resummed any non-perturbative contributions.
This does seem to help the link to asymptotic behaviour.

\mysection{The approach of L\"uscher et al.}%
L\"uscher, Sommer, Weisz and Wolff implemented a recursive strategy
which involves measuring a suitable observable on lattices of
different sizes, allowing them to calculate the running coupling with
no {\it a priori} assumptions. They have results for both
SU(2)~\rref\sutwofp{M.~L\"uscher, R.~Sommer, U.~Wolff, P.~Weisz Nucl.\
Phys.\ B389 (1993) 247}, and SU(3)~\rref\suthreefp{M.~L\"uscher,
R.~Sommer, P.~Weisz, U.~Wolff, ``A precise determination of the
running coupling in the SU(3) Yang-Mills theory'', preprint
hep-lat/9309005 (DESY 93-114)}.

\midinsert
\vskip 4in
\includegraphics{lusch2.eps}
\medskip
\leftline{\bigbold Figure 5\stepone: The strategy of L\"uscher,
Sommer, Weisz and Wolff.}
\endinsert

The idea (see figure 5) is that one performs simulations on two
lattices at a fixed coupling, so that one can change the size of the
lattice (used to set the scale) by an exact factor simply by adding
more points in each direction.  By inspired guesswork one can then
find a lattice at a different (higher) coupling which is roughly the
same physical size; any discrepancy can be handled with only small
errors by the renormalisation group.  Hence the size can be increased
stepwise.  The smallest lattice used is well in the perturbative
region: a small box has small length scales which means it is
perturbative by virtue of asymptotic freedom.  Note that the choice of
the measured observable is important: the response to a
chromoelectric field forced onto the lattice by the boundary
conditions is used.

This technique avoids our problem of having all the scales on the one
lattice, so that having a finite cut-off is less of a problem.  The
obvious point against is that there are more technical details to
understand and bring under control (for example, a lattice
renormalisation of the chosen observable used to extract the running
coupling).

The measured running couplings agree much better with na\"\i ve
perturbation theory than the UKQCD results in both SU(2) and SU(3).
The reasons are not understood; it may simply be (as L\"uscher et al.\
observe~\oref\suthreefp) that their chosen observable has particularly
good asymptotic scaling properties.

\mysection{Summary}%
As a summary of the results in SU(2) (as these are computationally
easier to extract; those in SU(3) look very similar), here is a table
of various estimates for the ratio of lattice spacings between two
inverse couplings $\beta=2.7$ and 2.85.  This deliberately uses the
bare coupling, except where explicitly improved, to show the
discrepancies; as explained above this is not generally the smartest
thing to do.  They range from the UKQCD results, through our results
improved by $\beta_E$, the results of ref.~\oref\sutwofp\ (which can
be read off from a graph in that paper using the authors' own fit), to
the na\"\i ve perturbative result at two loops.  Apart from the UKQCD
result (which includes an estimate of systematic errors), the errors
are negligible.

\bigskip

\def\sbox#1{\vtop{\hsize 5in #1}}
\settabs 4 \columns \def\thing#1{\hbox to 0pt{\hss #1 \hss}}

{\raggedright \pretolerance=10000 \parindent 0pt \parskip 5pt
\+{\stepone Quantity}&&& $a(2.7)/a(2.85)$\cr
\smallskip
\+\sbox{UKQCD:  Full fit to the potential, including
systematic errors}&&&1.60(6)\cr
\+\sbox{Perturbation theory with
first-order $\beta_E$ improvement}&&&1.53\cr
\+\sbox{Ref.~\oref\sutwofp}&&&1.48\cr
\+\sbox{Ordinary perturbation theory at two loops}&&&1.46\cr
}

\mysection{Conclusions}%
Attempts to see the running coupling in lattice gauge theory and to
relate it to perturbation theory have been successful; the necessity
of dealing with the problems of perturbation theory on the lattice is
now clear.  By setting a physical scale $\LQCD$ can be extracted.  As
time goes by we hope to have more control over the systematic errors
involved.  It is already claimed by analysts of the heavy quark
data~\oref\leplat\ that the lattice is a better place to extract the
running coupling than experiment; lattice theorists will need to
substantiate this by continuing to refine their methods.

\references

\bye